\documentclass{article}
\usepackage{bbding}
\usepackage{makecell}
\usepackage{amsfonts}

\usepackage{ijcai24}

\usepackage{times}
\usepackage{soul}
\usepackage{url}
\usepackage[hidelinks]{hyperref}
\usepackage[utf8]{inputenc}
\usepackage[small]{caption}
\usepackage{graphicx}
\usepackage{amsmath}
\usepackage{amsthm}
\usepackage{booktabs}
\usepackage{algorithm}
\usepackage{algorithmic}
\usepackage[switch]{lineno}
\usepackage{graphicx}
\usepackage{subcaption}
\usepackage{color}
\usepackage{colortbl}
\usepackage{xcolor}
\usepackage{multirow}

\title{Speech-Forensics: Towards Comprehensive Synthetic Speech Dataset Establishment and Analysis}

\author{
Zhoulin Ji$^1$
 \and
Chenhao Lin\footnotemark[1]$^2$ \and
Hang Wang$^{3,4}$ \And
Chao Shen$^2$\\
\affiliations
$^1$School of Software Engineering, Xi'an Jiaotong University\\
$^2$School of Cyber Science and Engineering, Xi'an Jiaotong University\\
$^3$School of Automation Science and Engineering, Xi’an Jiaotong University\\
$^4$Department of Computing, The Hong Kong Polytechnic University\\
\emails
310449@stu.xjtu.edu.cn,
\{linchenhao, cshangwang\}@xjtu.edu.cn,
 chaoshen@mail.xjtu.edu.cn
}

\begin{document}

\maketitle
\footnotetext[1]{Corresponding authors}

\begin{abstract}
Detecting synthetic from real speech is increasingly crucial due to the risks of misinformation and identity impersonation. While various datasets for synthetic speech analysis have been developed, they often focus on specific areas, limiting their utility for comprehensive research. To fill this gap, we propose the Speech-Forensics dataset by extensively covering authentic, synthetic, and partially forged speech samples that include multiple segments synthesized by different high-quality algorithms. Moreover, we propose a \textbf{TE}mporal \textbf{S}peech Localiza\textbf{T}ion network, called TEST, aiming at simultaneously performing authenticity detection, multiple fake segments localization, and synthesis algorithms recognition, without any complex post-processing. TEST effectively integrates LSTM and Transformer to extract more powerful temporal speech representations and utilizes dense prediction on multi-scale pyramid features to estimate the synthetic spans. Our model achieves an average mAP of 83.55\% and an EER of 5.25\% at the utterance level. At the segment level, it attains an EER of 1.07\% and a 92.19\% F1 score. These results highlight the model's robust capability for a comprehensive analysis of synthetic speech, offering a promising avenue for future research and practical applications in this field.
\end{abstract}

\section{Introduction}
Speech forensics analysis constitutes a crucial component in the field of forensic science, focusing on the verification of the authenticity and integrity of digital audio recordings. With the advent of advanced artificial intelligence, deep learning technologies have significantly progressed, particularly in the realm of content generation with notable examples including WaveNet~\cite{oord2016wavenet}, Tacotron~\cite{wang2017tacotron}, or VITS~\cite{kim2021conditional}. The rapid evolution of text-to-speech (TTS) synthesis and voice conversion (VS) technology has markedly narrowed the perceptual gap between deepfake audio and authentic audio recordings, making them virtually indistinguishable from the human ear. This development poses substantial challenges in the domain of speech forensics, as forensic experts increasingly face difficulties in ascertaining the authenticity and integrity of audio recordings, which are often critical pieces of evidence. 

Extensive research has been conducted to address the forensic challenges posed by deepfake speech, and several benchmarks have been established. The ASVspoof databases~\cite{wang2020asvspoof,yamagishi2021asvspoof} stand out as a pivotal resource, designed to bolster research in developing utterance-level countermeasures against deceptions in automatic speaker verification systems. ADD~\cite{yi2022add,yi2023add} represents another significant dataset series for detecting fake audio, encompassing a range of detection tasks under various scenarios, which includes the localization of tampered regions in partially forged speech and the recognition of utterance-level forgery algorithms. Simultaneously, a diverse range of studies have been proposed to navigate the intricate scenarios of speech analysis in practical settings. Initially, the research relied on identifying artifacts to verify the authenticity and integrity of audio recordings~\cite{zhao2013audio}. Traditional methods predominantly focused on the acoustic and statistical disparities between forged and genuine audio to facilitate differentiation~\cite{kajstura2005application,yang2008detecting,malik2010audio}. As the complexity of forensic scenarios escalated, machine learning, especially deep learning techniques, began to receive heightened attention from researchers, leading to significant breakthroughs in numerous instances~\cite{villalba2015spoofing,alzantot2019deep,li2023multi,zeng2023deepfake}.

While previous datasets in this field offer valuable insights, they also come with their own set of limitations and focuses. The majority of ASVspoof datasets, for instance, are primarily geared towards binary classification at the utterance level. The ADD database, although it encompasses scenarios of partially forged speech, tends to oversimplify these scenarios by typically involving alterations in just a single region of the speech. The PartialSpoof dataset~\cite{zhang2021initial} has made strides by updating its segmentation labels to enable the localization of multiple forged regions within a speech sample. However, this dataset does not provide corresponding labels for the forgery algorithms used, which restricts our ability to glean vital forensic insights regarding the origins of the forgery. Consequently, the algorithms \cite{tak2021end,cai2023dku,lu2023detecting} developed based on these datasets often analyze synthesized audio from a singular perspective. This approach substantially limits their broader applicability and impedes the progression toward a more comprehensive and multi-dimensional forensic analysis.

To tackle these challenges, we develop a sophisticated data pipeline for creating an extensive synthetic speech analysis dataset, which we have named Speech-Forensics. This dataset is meticulously crafted to maintain the semantic integrity of each audio sample. We achieve this by randomly manipulating multiple regions within a single audio and meticulously recording the specific forgery algorithms applied to these regions. This approach enables our dataset to support a thorough evaluation of speech analysis algorithms, encompassing aspects such as authenticity detection, regional localization, and algorithmic recognition. 

In addition, we introduce the TEmporal Speech localizaTion network called TEST. With a strong emphasis on temporal localization, this framework facilitates the seamless execution of complex tasks, including authenticity detection, regional localization, and forgery algorithm recognition, in a single and streamlined process. TEST negates the need for intricate post-processing steps, thereby allowing for a more flexible and efficient analysis of diverse synthesized speech forgeries. Drawing inspiration from the realm of temporal action localization, TEST utilizes a one-dimensional feature pyramid network, along with the combination of masked difference convolutions, LSTM, and transformers, enabling it to perform dense prediction tasks effectively. To our knowledge, this is the first instance of such a comprehensive dataset and framework being implemented in the field of synthetic speech analysis. The dataset and code are available at \url{https://github.com/ring-zl/Speech-Forensics}. Our contributions can be summarized as follows:

\begin{itemize}
    \item We implement a rational data pipeline and establish a comprehensive synthetic speech analysis dataset, encompassing multi-regional forgeries and their corresponding algorithms.
    \item We propose a framework designed for simultaneous authenticity detection, localization of multiple fake segments, and recognition of synthesis algorithms, without any complex post-processing.
    \item We demonstrate the robust capability of our proposed framework to perform comprehensive, multi-dimensional analyses of synthetic speech.
\end{itemize}

\begin{table*}[t]
    \centering
    \begin{tabular}{l|c|c|c|c|c|c|c}
        \hline
       \textbf{Datasets} & \textbf{Year} & \textbf{Language}  & \textbf{\#Fake Types}  & \textbf{\#Det.} & \textbf{\#Loc.} & \textbf{\#Reg.}   & \textbf{\makecell[c]{\#Spans\\per Sample}}  \\
        \hline 
        ASVspoof 2019~\cite{wang2020asvspoof} & 2019 &English      & TTS/VC     & \Checkmark & \XSolidBrush & \Checkmark    & Fully                             \\
        FoR~\cite{8906599}          & 2019 &English      & TTS        & \Checkmark & \XSolidBrush & \XSolidBrush  & Fully                              \\  
        In-the-Wild~\cite{muller2022does}  & 2021 &English       & TTS        & \Checkmark & \XSolidBrush & \XSolidBrush  &Fully                              \\  
        PartialSpoof~\cite{zhang2021initial} & 2021 &English       & TTS/VC     & \Checkmark & \Checkmark   & \XSolidBrush  & Multiple                           \\
        ASVspoof 2021~\cite{yamagishi2021asvspoof} & 2021 &English    & TTS/VC     & \Checkmark & \XSolidBrush & \Checkmark    & Fully                            \\
        WaveFake~\cite{frank2021wavefake}      & 2022 &English      & TTS        & \Checkmark & \XSolidBrush & \Checkmark    & Fully                            \\   
        SASV 2022~\cite{jung2022sasv}     & 2022 &English      & TTS/VC     & \Checkmark & \XSolidBrush & \Checkmark    & Fully                           \\
        ADD 2022~\cite{yi2022add}      & 2022 &Chinese      & TTS/VC     & \Checkmark & \XSolidBrush & \XSolidBrush  & Fully                           \\
        ADD 2023 Track 2~\cite{yi2023add}      & 2023 &Chinese & TTS/VC     & \Checkmark & \Checkmark   & \XSolidBrush  & Single                          \\
        ADD 2023 Track 3~\cite{yi2023add}      & 2023 &Chinese & TTS/VC     & \Checkmark & \XSolidBrush & \Checkmark    & Fully                         \\

        \rowcolor{gray!30}
        Speech-Forensics & 2024 &English  & TTS+VC     & \Checkmark & \Checkmark   & \Checkmark    & Multiple                        \\ 
        
        \hline
    \end{tabular}
    \caption{The statistics of existing synthetic speech datasets. For Fake Types, the samples in our dataset may contain both TTS and VC algorithms. “Det.”, “Loc.” and “Reg.” stand for authenticity detection, forgery localization, and synthesis algorithm recognition. The “Spans per Sample” indicates whether samples in the dataset are fully fake, partially fake with a single synthetic segment, or composed of multiple synthetic segments.}
    \label{tab: existing Datasets}
\end{table*}

\section{Related Work}
\subsection{Existing Synthetic Speech Datasets}
Table~\ref{tab: existing Datasets} summarizes key datasets of synthetic speech analysis, illustrating the evolution and focus of current research. The inaugural ASVspoof Challenge in 2015, based on the SAS database~\cite{wu2015sas}, concentrated on detecting spoofed speech. ASVspoof 2019~\cite{wang2020asvspoof} marked a significant development by introducing VC and TTS algorithms, establishing a critical baseline dataset for synthetic speech detection. In 2020, Reimao employed deep-learning synthesizers to create the FoR~\cite{8906599} dataset. In 2021, the In-the-Wild dataset~\cite{muller2022does} emerged as the first to represent real-world scenarios in the field. In the same year, the Partialspoof dataset~\cite{zhang2021initial} was introduced, becoming the first dataset to propose the concept of partial forgery. It established the groundwork for the localization of forged regions, setting a precedent in the field. The ASVspoof 2021 challenge~\cite{yamagishi2021asvspoof} added a track for detecting synthetic and compressed tampered audio to enhance system robustness. The WaveFake~\cite{frank2021wavefake} dataset employs a broader range of advanced synthesis algorithms, significantly increasing the challenge of distinguishing forged audio. Building upon these developments, SASV Challenge 2022~\cite{jung2022sasv} concentrated on enhancing both Countermeasure (CM) and Automatic Speaker Verification (ASV) systems, aiming to achieve dependable detection of spoofed speech.

The ADD series, notably the ADD2022 challenge~\cite{yi2022add}, introduced tasks such as Low-quality and Partially Fake Audio Detection, and the Audio Fake Game, broadening the scope of forged audio analysis. ADD2023~\cite{yi2023add} further expanded this with tasks like Manipulation Region Location and Deepfake Algorithm Recognition, enabling a more thorough analysis of synthetic speech.

While these datasets provide a diverse landscape for spoofed audio analysis, each has its own focus and limitations. The ASV series, for instance, focuses on utterance-level analysis, whereas the ADD series primarily addresses single-segment localization. Although the PartialSpoof dataset has made strides, it lacks information on the spoofing algorithms used. This diversity in focus and the lack of comprehensiveness in existing datasets present challenges in conducting holistic forensic research in speech analysis.

\subsection{Synthetic Speech Analysis}
\paragraph{Detection.} Forged speech detection has advanced significantly, initially relying on various speech signal processing algorithms. Malik~\shortcite{malik2010audio} utilized spectral decay phenomena for forensic audio analysis. In early deceptive audio detection, machine learning techniques, such as SVM, were prevalent, with Alegre~\shortcite{alegre2012spoofing} proposing an SVM-based robust voice spoofing detection strategy. The advent of deep learning brought about extensive use of these algorithms in anti-spoofing efforts. For instance, LCNN~\cite{lavrentyeva2019stc} and RawNet2 serve as baseline models in the ASVspoof competition~\cite{todisco2019asvspoof}, achieved top performance in single-system detection.

\paragraph{Localization.} Research has progressed beyond utterance-level speech forgery detection to address the more complex task of localizing specific forgery regions in partially fake speech. Li~\shortcite{li2023convolutional} viewed audio as a sequence of frames, using a CRNN network for frame classification. However, this indirect localization method often results in less intuitive audio analysis outcomes. Cai~\shortcite{cai2023dku} introduced a boundary prediction-based model focusing on clipped segment boundaries, though it depends on dataset specifications and a sophisticated post-processing procedure. Despite these advancements, a simple, intuitive, and flexible method for accurate localization remains a pressing need in the field.

\paragraph{Recognition.} With the increasing use of deepfake algorithms for speech synthesis, the field has moved beyond basic authenticity classification to include detailed forgery information analysis, like identifying specific generating algorithms. Qin~\shortcite{qin2022speaker} drew parallels between deepfake algorithm recognition and speaker verification, achieving notable results. To address the challenge of unknown algorithms, Zeng~\shortcite{zeng2023deepfake} extended their closed-set experiments using the ECAPA-TDNN model and data augmentation, successfully generalizing to new algorithms. However, most existing work focuses on utterance-level recognition, leaving the challenge of identifying multiple forgery algorithms within a single audio sample largely unexplored.


\begin{figure}[t]
  \centering
  \begin{subfigure}[b]{0.15\textwidth}
    \includegraphics[width=\textwidth]{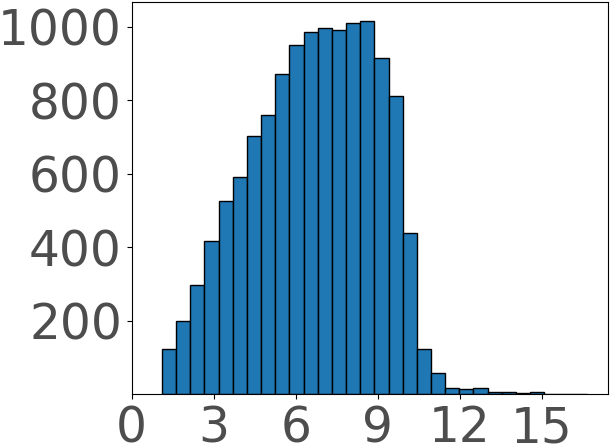}
    \caption{Duration Time}
    \label{fig:sub1}
  \end{subfigure}
  \hfill
  \begin{subfigure}[b]{0.15\textwidth}
    \includegraphics[width=\textwidth]{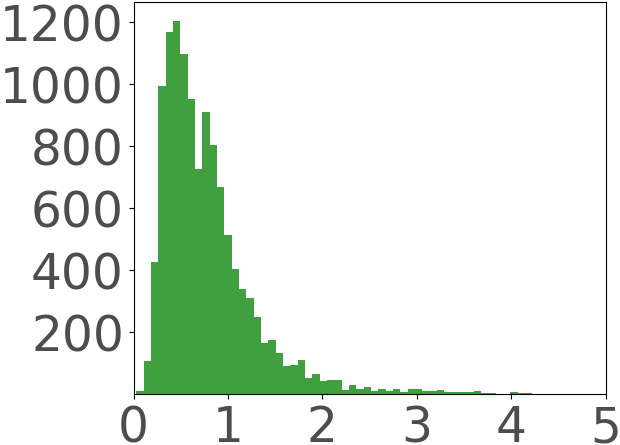}
    \caption{Span Length}
    \label{fig:sub2}
  \end{subfigure}
  \hfill
  \begin{subfigure}[b]{0.15\textwidth}
    \includegraphics[width=\textwidth]{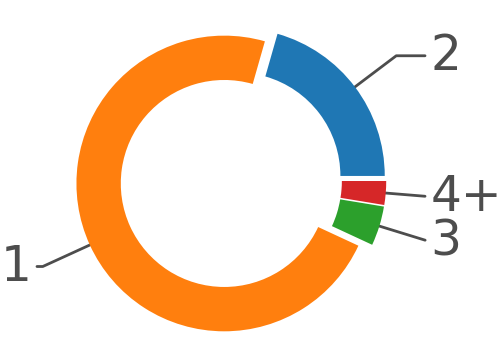}
    \caption{Algorithm Count}
    \label{fig:sub3}
  \end{subfigure}
    \caption{Distribution of the Speech-Forensics Dataset: (a) Number of samples (y-axis) across various durations (x-axis); (b) Frequency of fake spans (y-axis) across various lengths (x-axis); (c) Proportion of samples utilizing different numbers of synthesis algorithms.}

  \label{fig:total}
\end{figure}

\section{Speech-Forensics Dataset}
The research community has developed various datasets to tackle synthetic speech analysis tasks, each contributing uniquely to the field. However, comprehensive forensic analysis, especially in scenarios requiring multifaceted evidence evaluation, demands a dataset that transcends conventional limitations. Our dataset, Speech-Forensics, is designed to address this need. It aims to facilitate studies within a comprehensive analytical framework, catering to the nuanced requirements of forensic speech analysis. Speech-Forensics distinguishes itself by offering a diverse range of speech samples, encompassing various types of forgeries, and is structured to support both broad and detailed analyses.

\subsection{Dataset Construction}
To emulate the intricacies of real-world voice forensics, our dataset construction prioritizes semantic coherence in the fabricated samples. Based on insights from previous studies~\cite{yi2021half}, we employed a multi-step data pipeline consisting of: (1) Content Selection: Choosing appropriate content for editing. (2) Audio Synthesis: Generating audio from the edited text. (3) Audio-Text Alignment: Creating alignments between the audio and text. (4) Forgery Splicing: Using these alignments for splicing in speech forgeries.

We utilized the LJ Speech dataset\footnote{\url{https://keithito.com/LJ-Speech-Dataset/}}, a public domain collection of 13,100 audio clips with transcripts, read by a single speaker. This dataset, spanning approximately 24 hours, was chosen for its diversity in content and audio length (1 to 10 seconds per clip). To ensure semantically coherent modifications, we implemented a keyword replacement strategy. This involved using a Named Entity Recognition (NER) algorithm to identify and replace various entities (e.g., persons, organizations) within the texts. To diversify our entity pool, additional entities were extracted from a news dataset\footnote{\url{https://www.kaggle.com/datasets/therohk/million-headlines}}. Table \ref{tab: pool} details the distribution of these entities. Moreover, we integrated an antonym replacement strategy for adjectives in the transcripts, randomly substituting them with their antonyms to mirror the varied scenarios of voice forgery encountered in practical applications.

\begin{table}[t]
    \centering
    \begin{tabular}{lcccccc}
        \toprule
        \textbf{Set} & \textbf{\#Tim} & \textbf{\#Dat} & \textbf{\#Loc} & \textbf{\#Mon} & \textbf{\#Org} & \textbf{\#Psn} \\
        \midrule
        LJ   & 14  & 174  & 388  & 28 & 798 & 975 \\
        News & 222 & 5490 & 3016 & 488 & 3367 & 2384 \\
        \bottomrule
    \end{tabular}
    \caption{The entity pool is constructed by two datasets, where the abbreviations in the table respectively represent time, date, location, money, organization, and person.}
    \label{tab: pool}
\end{table}

To enhance the naturalness of the fabricated samples, we avoid directly synthesizing individual keyword segments. Instead, we generate complete synthetic audio for the entire edited transcript. Then we extract the specifically forged keyword phrases for final assembly. This method ensures a seamless integration of the forged segments with the original audio context. In selecting algorithms, we prioritized models pre-trained on the LJ Speech dataset. This choice aims to retain as much of the original speaker's characteristics as possible, thereby preserving the authenticity of the synthesized audio. The specific configurations of the TTS and VC algorithms used in our dataset construction are detailed in Table \ref{tab: TTS/VC}.

\begin{table}[t]
    \centering
    \begin{tabular}{l|l|c}
    \toprule
    \textbf{Category} & \textbf{Name}      & \textbf{Modeling} \\
    \midrule
    \multirow{5}{*}{Acoustic Model} & Tacotron-DDC-ph & Seq2Seq \\
                                    & Tacotron2-DCA   & Seq2Seq \\
                                    & Neural-HMM      & Seq2Seq \\
                                    & Glow-TTS        & Flow \\
                                    & OverFlow        & Flow \\
    \midrule
                                    
    \multirow{3}{*}{Vocoder}        & Multiband-MelGAN & GAN \\
                                    & HiFiGAN-v2      & GAN \\
                                    & UnivNet         & GAN \\
    \midrule
    \multirow{3}{*}{End-to-End}     & Your-TTS        & CVAE \\
                                    & FreeVC          & VAE \\
                                    & VITS            & CVAE \\

    \bottomrule
    \end{tabular}
    \caption{Basic Information of Synthesis Algorithms.}
    \label{tab: TTS/VC}
\end{table}

Upon obtaining the original audio and its corresponding transcription, along with the series of synthesized audio files and their edited transcripts, we employed the Montreal Forced Aligner (MFA) tool. This tool utilizes a trained speech recognition system to produce accurate timestamp alignments between the audio and text. These timestamps are instrumental in seamlessly integrating the synthesized segments into the original audio. This meticulous process results in the creation of realistic voice forgeries, which are accurately labeled and closely mirror real-world scenarios. The integration strategy ensures that the fabricated samples are not only natural-sounding but also maintain coherent and reasonable semantics, thus enhancing their applicability in forensic analysis.

\begin{figure*}[t]
    \centering
    \includegraphics[width=0.95\linewidth]{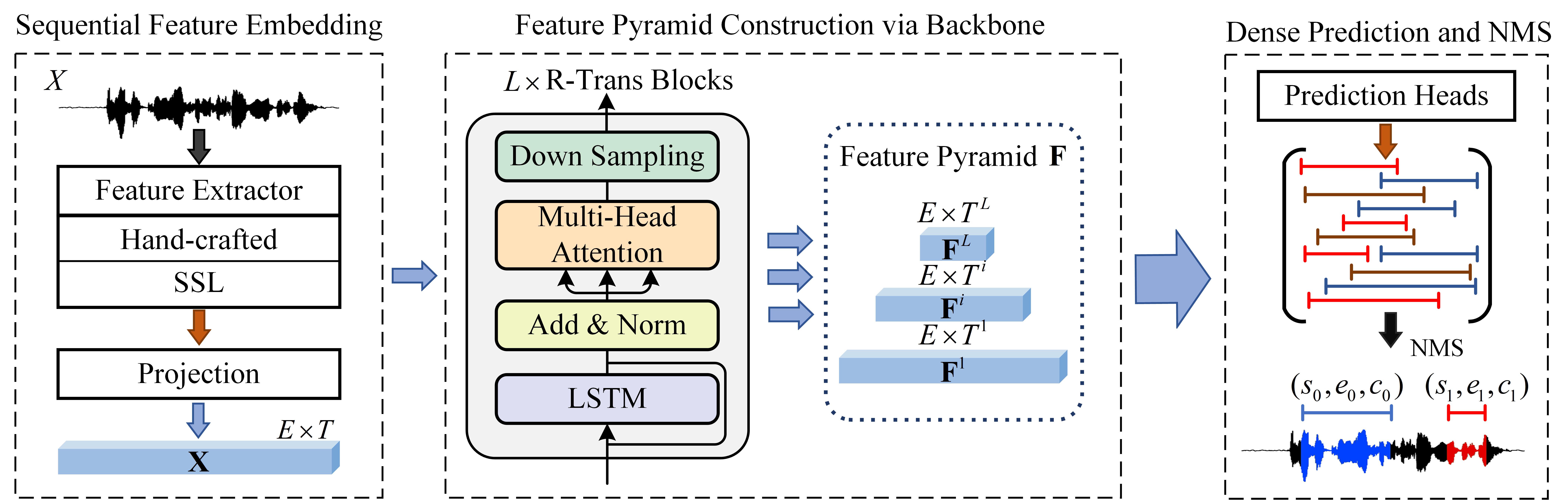}
    \caption{Overview of our method TEST. We process the given speech samples \(X\) through an embedding module to transform them into sequential features \(\mathbf{X}\). These features are then processed through a group of \(L\) R-transformer blocks to obtain multi-scale temporal feature representations, forming a temporal feature pyramid \(\mathbf{F}\). Two sets of prediction heads are used to predict the proposed forged speech segments and their corresponding forgery algorithms. Finally, the final analysis results are obtained through Non-Maximum Suppression.}
    \label{fig:method.png}
\end{figure*}

\subsection{Dataset Description}
\paragraph{Task Comprehensiveness.} Unlike existing datasets that typically analyze synthesized audio from a single perspective, our dataset provides comprehensive Ground Truth data. This includes the veracity of speech at the utterance level, timestamps for each forged span, and the specific synthesis algorithms used. This allows for more nuanced tasks in forensics.

\paragraph{Dataset Size.} We have extensively utilized the LJ Speech dataset, generating 4,323 records using NER and 3,129 records through antonym replacement strategies. To create a more realistic forensic environment, the remaining samples are labeled as bonafide. Figure~\ref{fig:sub1} illustrates the distribution of sample durations in the Speech-Forensics dataset.

\paragraph{Multiple Spans Per Speech.} Enhancing the dataset's utility, we manipulated multiple spans within each audio sample using different synthesis algorithms. This detailed approach provides granular timestamp information and corresponding algorithm data. The length distribution of forged spans within the dataset is presented in Figure~\ref{fig:sub2}.

\paragraph{Appropriate Algorithm.} A variety of advanced voice synthesis algorithms were selected for creating the voice forgeries. These algorithms, known for their natural-sounding synthetic outputs, were mostly trained on the LJ Speech dataset to mimic intricate forgery scenarios. The dataset includes both TTS and VC algorithms, with the number of algorithms used per sample depicted in Figure~\ref{fig:sub3}.

\section{Method}

\subsection{Problem Definition}
In forensic scenarios involving synthetic speech, conducting a multifaceted analysis is essential. This analysis must extend beyond merely detecting authenticity to include localizing tampered regions and identifying the methods used for forgery. Traditional localization methods, which focus only on identifying the true and false boundaries, prove inadequate for handling multiple forged intervals. These methods, along with frame-by-frame classification techniques, fall short as they do not offer a clear representation of the extent of forgeries and often require complex post-processing. Similarly, in the realm of algorithm recognition, most existing methods operate at the utterance level, presenting challenges in cases where speech forgeries involve multiple forging algorithms. This limitation hinders the complete identification of manipulations. To overcome the limitations of existing methods and enhance practicality, we have developed a comprehensive framework for synthesized speech analysis. Our approach diverges from existing methods by defining the task objective as a structured output: 

\begin{equation}
X \rightarrow  Y = \begin{cases} 
\emptyset, & \text{if } N = 0, \\
\{ y_1, y_2, \ldots, y_N \}, & \text{if } N > 0.
\end{cases}
\end{equation}

\noindent where \( X \) represents a potential synthetic speech, \( Y \) is our target for forensic speech analysis. \(N\) denotes the total number of forged spans in the entire audio. When \( N = 0\), it indicates that the audio is bonafide. Each span represented by \( y_i = ( s_i, e_i, c_i) \), which is defined by its starting time \( s_i \), ending time
\( e_i \) and the class of the algorithm used \( c_i \). It is evident that \( s_i \in [0, T] \) and \( e_i \in [0, T] \), where \( T \) denotes the total duration of the audio, and that \( s_i < e_i \).

To achieve this structured output objective, we were inspired by an anchor-free representation~\cite{yang2020revisiting,zhang2022actionformer}. The core idea is the dense regression of the current timestamp relative to the start and end distances of the forged span. This implies that we have transformed the output objective of the model into the following form:

\begin{equation}
X \rightarrow \hat{Y} = \{\hat{y_1}, \hat{y_2}, \ldots, \hat{y_M}\}
\end{equation}

\noindent where \(M\gg N\), \( \hat{y_\tau} = ( d_{\tau}^{s}, d_{\tau}^{e}, p(c_\tau) ) \) represents a proposal at a hypothetical timestamp \( \tau \) in our model, defined as
\(d_{\tau}^{s}\) and  \(d_{\tau}^{e} \) are the distances relative to the forged span's onset and offset. \( p(c_\tau) \) indicating the probability of the synthetic algorithm's category.

This dense prediction approach not only effectively utilizes the temporal information of speech but also facilitates comprehensive synthetic speech analysis. Inspired by temporal action localization~\cite{zhang2022actionformer}, we propose a Temporal Speech Localization Network as illustrated in Figure \ref{fig:method.png}, comprising an embedding module, a backbone, and a group of prediction heads. The subsequent sections provide detailed explanations of these components.

\subsection{Embedding Module}
Due to the sparsity of speech data (typically sampled at 16,000 points per second), researchers often opt to filter the original speech data~\cite{tak2021end}, extract spectrograms~\cite{li2023convolutional}, or employ deep features~\cite{cai2023dku} for experiments to achieve a more robust and effective representation of speech. 

For a given speech \( X \) of duration \(D\), it is transformed into a sequence of features \( \mathbf{X} = \{\mathbf{x}_1,\mathbf{x}_2,\ldots, \mathbf{x}_{T}\}\) via the embedding module. Here, each \( \mathbf{x}_i\) has \( E \) embedding dimensions and can represent a time segment of frame length \( f_L \). The temporal offset between each \( \mathbf{x}_i \) is defined as the frameshift \( f_S \), the length of the sequence \( T\) is given by the following equation:

\begin{equation}
T = \left\lfloor \frac{D - f_L}{f_S} \right\rfloor + 1
\end{equation}
\noindent where \(\lfloor * \rfloor\) indicates round down operation.

Our embedding module consists of a feature extractor and a projection layer. The feature extractor can either perform mathematical operations for extracting handcrafted features, or it can be a trained self-supervised learning network that extracts deep features from speech signals. In this work, we have employed both approaches. We use Linear Frequency Cepstral Coefficients (LFCC) and Mel-Frequency Cepstral Coefficients (MFCC) for their proven efficacy in capturing speech characteristics. Additionally, we leverage models like Wav2Vec2~\cite{baevski2020wav2vec} and WavLM~\cite{chen2022wavlm}, trained on diverse multilingual and unlabeled language data, to acquire high-level speech representations. Rather than fine-tuning these models for a specific task, we extract features from various layers. This method allows the model to benefit from a range of speech features, potentially enhancing task performance, as previous studies~\cite{yang2021superb} have shown.

The projection layer is another component, consisting of a one-dimensional masked difference convolution. This design is intended to not only adjust the input dimensions for the backbone but also to enrich the temporal information captured from the speech data. Traditionally, in handcrafted feature extraction, derivatives like first and second-order differences are included as supplementary parameters to cepstral coefficients. Our projection layer seeks to emulate this enriching process by using the differences convolution, a concept inspired by a temporal enhanced network~\cite{yu2021searching}. For a standard one-dimensional difference convolution with a mask, the feature output can be expressed as: 

\begin{equation}
\begin{aligned}
MDC(t_0) = & \sum_{t_n \in \mathcal{D}} w(t_n)\!\cdot\! x(t_0+t_n)  \\
           & + \theta \!\cdot\! (-x(t_0) \cdot \sum_{t_n \in \mathcal{D}} w(t_n))
\end{aligned}
\end{equation}

\noindent where \( t_0 \) denotes current timestamp while \( t_n \) enumerates the timestamps in 
\(\mathcal{D}\), \( w \) represents learnable weights, hyperparameter \( \theta \in [0,1] \) tradeoffs the contribution between intensity-level and gradient-level information.

\subsection{Backbone}
The backbone of our model, as illustrated in Figure~\ref{fig:method.png}, is composed of \(L\) R-Transformer blocks. These blocks incorporate the LSTM module into Transformer architecture, aimed at achieving more robust and powerful speech representations~\cite{sun2021transformer}. Drawing inspiration from hierarchical feature extraction methodologies~\cite{lin2017single,long2019gaussian}, we segment speech into various levels, facilitating the capture of features across different time scales.

This integration of R-Transformers with one-dimensional downsampling convolutions forms the Bottom-up process of our feature map construction. For the Top-down process, we employ nearest-neighbor upsampling, complemented by Lateral connections, resulting in a layered feature pyramid \(\mathbf{F}\) with \(L\) levels, denoted as \(\mathbf{F}=\{  \mathbf{F}^1,\mathbf{F}^2,\ldots, \mathbf{F}^L\}\). 

Given a virtual timestamp \(\tau\) at level \(i\) of the feature pyramid, and considering its cumulative scaling stride \(s_i\), we can map this virtual timestamp back to the corresponding real timestamp \(t\) in the original speech sequence. This mapping is defined by the equation:

\begin{equation}
t = \lfloor s_i / 2 \rfloor + \tau \cdot s_i
\end{equation}
\noindent where \(\lfloor * \rfloor\) indicates round down operation.

\subsection{Prediction Head}
The role of our prediction head is to decode the feature pyramid \(\mathbf{F}\), generated by the backbone, into our desired output \(\hat{Y}\). This decoding involves two primary components of one-dimensional convolutional networks: the classification head and the localization head.

\paragraph{Classification Head.}
For each virtual timestamp \(\tau\) on the feature pyramid \(\mathbf{F}\), the classification head is designed to output a probability vector of length \( C \), where \( C \) represents the maximum number of synthetic algorithm categories. This vector reflects the probability \( p(c_\tau) \) of timestamp \( \tau \) belonging to each category. The parameters for this process are consistently applied across all levels of the pyramid.

\paragraph{Localization Head.}
The objective of our localization head is to determine the distance between the timestamp \( \tau \) and the start and end times of the forged interval, corresponding to onset \(d_{\tau}^{s}\) and offset \(d_{\tau}^{e}\). The output is valid only when timestamp \( \tau \) is within a forged interval. To ensure accurate distance estimation, we use the ReLU activation function.

The conversion of the virtual timestamp \( \tau \) to the real-time \( t \) in the corresponding audio is a critical step in our backbone design. For a given time moment \( t \) in the audio, we can determine the most likely forgery algorithm \(c_t\), and the start \(s_t\) and end \(e_t\) points of the forged spans using the following formula:

\begin{equation}
c_t = \arg\max p(c_t), \quad s_t = t - d_{s}^{t}, \quad e_t = t - d_{e}^{t}
\end{equation}

\subsection{Loss Function}
We employed a sigmoid focal loss to measure the loss of the predicted categories and a DIoU loss to evaluate the onset and offset.  To balance these components effectively, we utilize the following loss function:
\begin{equation}
\mathcal{L} = \sum_{t}( \lambda \mathcal{L}_{cls} + (1-\lambda) \mathbb{I}_{t} \mathcal{L}_{loc} ) / T_+
\end{equation}
\noindent where \(\lambda\) serves as a balancing ratio between the two functions. \(\mathbb{I}_{t}\) is an indicator function, determining whether the timestamp \( t \) falls within a specified forgery region. \( T_+ \) represents the total number of positive samples.

\section{Experiments}

\subsection{Embedding Settings}
In the embedding process, we employed MFCC and LFCC as the primary handcrafted features. We set the frame length to 25 ms and the frameshift to 20 ms for these features. A Fourier transform, utilizing 256 filters and 2,048 FFT, was applied to produce 256-dimensional features. To enrich our cepstral representation, we concatenated the first and second-order delta spectra, resulting in a 768-dimensional feature. For deep feature extraction, we used Wav2Vec2 and wavLM as representatives of self-supervised learning models. Under the BASE configuration, both models transform every 20 ms of audio into a 768-dimensional vector. Under the LARGE configuration, these models output a 1,024-dimensional vector for every 20 ms of audio. In the experiment, we sequentially selected six layers of varying depths for comparison.

\subsection{Implemental Details}
Our experiments were conducted on a NVIDIA 4090 GPU. We adopted mini-batch training using the AdamW~\cite{loshchilov2017decoupled} optimizer. The training included a warmup phase of 5 epochs, followed by a cosine decay schedule for the learning rate. The initial learning rate was set at $1e-3$, with a weight decay of $1e-3$. We employed Non-Maximum Suppression (NMS) to effectively filter out redundant and less effective spans from our model's output.

\subsection{Evaluation Metrics}
To ensure a comprehensive assessment of our method's performance across different scenarios, we have employed a multifaceted approach to evaluation metrics.

Firstly, we utilize the average Mean Average Precision (mAP) across various temporal intersections over union (tIOU) thresholds to gauge the overall capability of our model in accurately locating forged regions and identifying the synthetic algorithm used. This metric effectively evaluates both the precision and recall of our model in a singular framework.

For utterance-level performance analysis, we adopted the Equal Error Rate (EER) metric, as used in ASVspoof 2019 LA. This threshold-independent measure reflects the point at which the false acceptance rate and false rejection rate are equal. Additionally, we extend this evaluation to segment-level analysis. Here, for each 0.01s audio segment, we assess our model's ability to accurately classify the segment as either bonafide or synthetic, offering a more granular view of performance.




This comprehensive set of evaluation metrics allows for a thorough assessment of our method's effectiveness in various aspects of synthetic speech analysis.

\subsection{Main Results}
In this section, we will demonstrate comprehensive capability in synthesizing speech of our model TEST, focusing on its performance in three key areas: authenticity detection, forged segment localization, and synthesis algorithm recognition. Additionally, we will examine the impact of the various configurations on performance.


\begin{table}[t]
    \centering
    \begin{tabular}{lccccc}
        \toprule
        \multirow{2}{*}[-0.75ex]{\textbf{Method}} & \multicolumn{2}{c}{\textbf{EER}} & \multirow{2}{*}[-0.75ex]{\textbf{F1}} & \multirow{2}{*}[-0.75ex]{\textbf{mAP}} \\
        \cmidrule(lr){2-3}
                       & Utterance & Segmental & & \\
        \midrule
        RawNet2        & 24.50  & -    & -   & - \\
        AASIST         & 27.56  & -    & -   & - \\
        WBD            & -      & 15.35 & 82.16 & - \\   
        \midrule
        \textbf{TEST}  & \textbf{5.25} & \textbf{1.07} & \textbf{92.19} & \textbf{83.55} \\
        \bottomrule
    \end{tabular}
    \caption{Comparison of Method (`-' Indicates Task Unsupported).}
    \label{tab:compare}
\end{table}

Table~\ref{tab:compare} reports the performance of various models on the Speech-Forensics dataset. We utilize utterance-level EER to reflect the models' performance in binary authenticity detection of partially forged audio. Segment-level EER and F1 scores are used to evaluate the models' capability in locating forged segments. Additionally, mAP is employed to comprehensively assess the models' proficiency in both localization and algorithm recognition. The RawNet2~\cite{tak2021end} and AASIST models~\cite{9747766}, which were originally designed to detect fully forged speech, exhibited a significant performance decline when confronted with partially forged audio. The WBD~\cite{10094774} model, employing a boundary modeling for audio, achieved better F1 scores in partial forgery detection. However, it still shows suboptimal performance in the more nuanced task of segment-level localization, as indicated by its segment-level EER.

Our model, TEST, not only exhibits the best overall performance but also demonstrates significant advantages across various metrics. These results indicate that TEST is capable of not only performing binary classification tasks efficiently but also excelling in accurately locating forged segments in multi-segment spoofed samples and effectively recognizing the algorithms used in their creation.

Additionally, it is noteworthy that the utterance-level EER is marginally higher than the segment-level EER. This may be attributed to the design of the model, where genuine speech is defined as completely negative samples, contributing less to the discriminative capabilities of our model.

\begin{table}[t]
    \centering
    \begin{tabular}{lcccc}
        \toprule
        \multirow{2}{*}[-0.75ex]{\textbf{Extractor}}  & \multicolumn{2}{c}{\textbf{EER}} & \multirow{2}{*}[-0.75ex]{\textbf{F1}} & \multirow{2}{*}[-0.75ex]{\textbf{mAP}} \\
        \cmidrule(lr){2-3}
                       & Utterance & Segmental & & \\
        \midrule
        LFCC           & 9.38     & 1.74     &84.30      &73.13  \\
        MFCC           & 8.25     & 1.57     &85.70    &75.35 \\
        Wav2Vec2-B  & 5.25     & 1.07   & 92.19   & 83.55 \\
        WaLM-B      & 4.75     & 0.90   &92.74    & 85.01  \\
        Wav2Vec2-L &1.49    & 0.95   &93.32       &90.36\\
        WaLM-L     & \textbf{0.64}     & \textbf{0.43}     & \textbf{96.44}      & \textbf{95.26}\\
        \bottomrule
    \end{tabular}
    \caption{Comparison of Feature Extractor.}
    \label{tab:embedding}
\end{table}

Table~\ref{tab:embedding} presents the impact of different feature extractors on model performance. Traditional handcrafted features such as MFCC and LFCC are noticeably less effective compared to deep learning-based features. Deep feature extractors, as demonstrated by the results, exhibit superior adaptability for analyzing synthetic speech. Enhancing and scaling up these deep features consistently leads to significant performance improvements. This shows that Wav2Vec and WaLM variants offer more nuanced and effective feature representation for complex tasks such as synthetic speech detection.

\begin{table}[t]
\centering
\begin{tabular}{ccc}
\toprule
\multirow{2}{*}[-0.75ex]{\textbf{Feature Map Depth}} & \multicolumn{2}{c}{\textbf{mAP}} \\
        \cmidrule(lr){2-3}
                       & Wav2vec2 & WavLM  \\ 
\midrule
Index 0     & 83.55       & 85.01        \\
Index 2     & \textbf{85.61}        & \textbf{88.84}        \\
Index 4     & 75.79       & 82.54        \\
Index 6     & 68.85       & 62.70         \\
Index 8     & 73.86        & 47.37        \\
Index 10     & 61.85       & 43.77        \\ \bottomrule
\end{tabular}
\caption{Comparison of mAP for SSL Model Across Depths.}
\label{tab:index}
\end{table}

Table~\ref{tab:index} illustrates the performance variations in mAP when using feature maps of different depths. Overall, a consistent observation across both Wav2Vec2 and WavLM models is that shallower features demonstrate stronger adaptability for comprehensive synthesized speech analysis. This trend may be attributed to the deeper features' development for semantic self-supervision, encapsulating richer semantic information. For the semantically coherent forged samples in the dataset, the semantic information does not significantly aid in distinguishing them. In contrast, shallower features offer greater flexibility and adaptability. Yet, excessively shallow features can become overly generalized, leading to suboptimal performance in detecting nuances in synthesized speech.

\section{Discussion}
We present a viable benchmark for addressing comprehensive synthetic speech analysis and propose an innovative solution approach. Nonetheless, the Speech-Forensics dataset is constrained by limited resources, requiring further expansion in both data volume and speaker diversity. Additionally, while we have introduced TEST as a foundational method for comprehensive speech analysis, further exploration into tailored research methodologies specific to this domain remains imperative. These aspects represent potential avenues for refinement in future research endeavors.

\section{Conclusion}
This study has developed an extensive synthetic speech analysis dataset, named Speech-Forensic. This dataset comprises authentic, synthetic, and various high-quality algorithm-generated, multiple-segment partially forged speeches, effectively bridging the gap in comprehensive datasets within the synthetic speech analysis field. Furthermore, we introduce TEST, a temporal speech localization network to effectively integrate difference convolution, LSTM, and transformers, leveraging multi-scale pyramid features for dense prediction. This method can simultaneously perform authenticity detection, localization of multiple forged segments, and synthesis algorithm identification, without requiring complex post-processing. Experimental results demonstrate that TEST exhibits robust capabilities across these comprehensive tasks and achieves exceptional performance in each sub-task.

\section*{Acknowledgements}
This work was supported in part by the National Key Research and Development Program of China under Grant 2021YFB3100700; the National Natural Science Foundation of China under Grant 62376210, 62161160337, 62132011, U21B2018, U20A20177, U20B2049; the Shaanxi Province Key Industry Innovation Program under Grant 2023-ZDLGY-38. Chenhao Lin is the corresponding author.

\bibliographystyle{named}
\bibliography{ijcai24}

\end{document}